\documentstyle{mn}
\input{epsf}

\newcommand{\etalMNRAS}{\mbox{et al. }}
\newcommand{\pmt}{\mbox{$\pm \;$}}
\newcommand{\rtp}[1]{\mbox{$^{#1}$}}

\def\lesssim{\mathrel{\hbox{\rlap{\hbox{\lower4pt\hbox{$\sim$}}}\hbox{$<$}}}}
\def\gtrsim{\mathrel{\hbox{\rlap{\hbox{\lower4pt\hbox{$\sim$}}}\hbox{$>$}}}}

\newcommand{\Ht}{\mbox{${\rm H_2}$}}
\newcommand{\HI}{\mbox{{\rm H \footnotesize{I} }}}
\newcommand{\HII}{\mbox{{\rm H \footnotesize{II} }}}
\newcommand{\RSUN}{\mbox{$R_0 \;$}}
\newcommand{\RSUNn}{\mbox{$R_0$}}
\newcommand{\VSUN}{\mbox{$\Theta_0 \;$}}
\newcommand{\VSUNn}{\mbox{$\Theta_0$}}

\newcommand{\OA}{\mbox{$A$}}
\newcommand{\OB}{\mbox{$B$}}
\newcommand{\OAR}{\mbox{$A(R)$}}
\newcommand{\OBR}{\mbox{$B(R)$}}
\newcommand{\hR}{\mbox{$h_{\rm R}$}}
\newcommand{\MSpcsq}{\mbox{$M_{\odot}\,{\rm pc}^{-2}$}}

\newcommand{\kms}{\mbox{${\rm km \;s}^{-1}$}}
\def\apj{ApJ}

\def\aj{AJ}
\def\mnras{MNRAS}
\def\aap{A\&A}
\def\aaps{A\&AS}

\newcommand{\SOTON}{Dept.  of Physics and Astronomy, \\
University of Southampton, Southampton SO17 1BJ, U.K.}

\begin{document}

\title[Refining the Oort and Galactic constants]
      {Refining the Oort and Galactic constants}

\author[R. P. Olling, M. R. Merrifield]
{       Rob P. Olling$^1$\thanks{E-mail: olling@astro.soton.ac.uk},
Michael R. Merrifield$^1$\thanks{E-mail: mm@astro.soton.ac.uk}    \\
$^1$\SOTON }

\date{Accepted 1998 February 2}

\maketitle


\begin{abstract}

The local stellar kinematics of the Milky Way offer a useful tool for
studying the rotation curve of the Galaxy.  These kinematics -- usually
parameterized by the Oort constants $A$ and $B$ -- depend on the local
gradient of the rotation curve as well as its absolute value
($\Theta_0$), and the Sun's distance to the Galactic center ($R_0$).  The
density of interstellar gas in the Milky Way is shown to vary
non-monotonically with radius, and so contributes significantly to the
local gradient of the rotation curve.  We have therefore calculated mass
models for the Milky Way that include this component, and have derived
the corresponding radial variation in the Oort constants.  Between
0.9\RSUN and 1.2\RSUN the Oort {\em functions} \OAR\ and \OBR\ differ
significantly from the general $\sim \Theta_0/R$ dependence.  Various
previously-inexplicable observations are shown to be consistent with
these new predictions.  For example, these models may explain the
$\sim$40\% difference between the values for $2 A \RSUN$ derived from
radial velocity data originating in the inner and outer Galaxy
\cite{mrM92}.  They also go some way toward explaining the different
shapes of the velocity ellipsoids of giant and dwarf stars in the solar
neighbourhood.  However, a consistent picture only emerges if one adopts
small values for the radius of the solar circle (\RSUN = 7.1 \pmt 0.4
kpc) and local circular speed (\VSUN = 184 \pmt 8 \kms).  With these
Galactic constants the Milky Way's rotation curve declines slowly in the
outer Galaxy; $V_{\rm rot}$(20 kpc) = 166 \kms.  Our low value for the
distance to the Galactic center agrees well with the only direct
determination of \RSUN (7.2 \pmt 0.7 kpc, Reid 1993).  Using these
Galactic constants, we find that the proper motion of Sgr A$^*$ is
consistent with the observational constraints \cite{dcBraS87,dcB96,mjR98}. 
Simple analytic arguments as well as detailed calculations show that the
radial velocities and proper motions of our best fit model are entirely
consistent with the radial velocities of Cepheids \cite{PMB94} and the
Hipparcos measurements of their proper motions \cite{mFpW97}.

\end{abstract}


\newpage

\vspace*{-0mm}
\section{Introduction}
\label{sec:Introduction}

Since the first measurements of the Milky Way's rotation almost
seventy years ago, a great deal of effort has been expended in
studying its kinematics \cite{EGmFsT91REF}.  The simplest kinematic
quantity that we can determine is the rotation curve, $\Theta(R)$,
which measures the speed of circular orbits as a function of radius.
Unfortunately, our location within the Milky Way means that even this
simple dynamical quantity is difficult to establish.  In particular,
the modest uncertainties in our knowledge of the distance to the
Galactic center (\RSUN = 7.7 \pmt 0.7 kpc; Reid 1993) and the rotation
speed at the solar circle (\VSUN = 200 \pmt 20 \kms; Sackett 1997)
lead to significant uncertainties in the inferred form for $\Theta(R)$
(Fich \& Tremaine 1991).  Measurements of stellar kinematics provide
one approach to constraining the form of the rotation curve.  The
quantities measured are parameterized by the Oort functions,

\begin{eqnarray}
A(R) &=&  +\frac{1}{2} 
     \left( \frac{\Theta(R)}{R} - \frac{d\Theta(R)}{dR} \right)
     \label{eqn:Oort_A} \\
B(R) &=& -\frac{1}{2} 
     \left( \frac{\Theta(R)}{R} + \frac{d\Theta(R)}{dR} \right).
     \label{eqn:Oort_B}
\end{eqnarray}

\noindent For a flat rotation curve we see that both $A$ and $B$ are
inversely proportional to $R$, with smaller $A$ and larger $B$ values in
the outer Galaxy.  The quantity $B$ is determined by studying the
dependence of stellar proper motions on galactic longitude, while $A$
can be calculated from analysis of either proper motions ($\mu_l$) or
radial velocities ($v_{r}$):

\begin{eqnarray}
v_{r} &=& A \ d \sin{2 l} 
      \label{eqn:Vrad_apx} \\
\mu_l &=& \frac{A \cos{2 l} + B}{4.74} \ ,
      \label{eqn:mu_l_apx}
\end{eqnarray}

\noindent with $d$ the distance to the object (in units of kpc), $l$
Galactic longitude, $A$ and $B$ in \kms\ kpc\rtp{-1}, and $\mu_l$ in
milli arcsec (mas) per year \cite{MB81}.  Note that the above equations
for $v_r$ and $\mu_l$ are only correct for $\frac{d}{R_0} << 1$.

It is also possible to determine $A-B = \Theta/R$ essentially
independently from the individual values of $A$ and $B$ from proper
motion surveys in the directions of $l=90$ degrees and $l=270$ degrees. 
In these directions, objects have a rather limited range in
Galactocentric radius, and are thus only slightly affected by the radial
dependence of the Oort functions. 

Finally, the combination $-B/(A-B)$ can be estimated from the shape of
the velocity ellipsoid of random stellar motions.  Although the Oort
{\em functions} vary with Galactic radius, stellar kinematic
observations have only determined the Oort $A$ and $B$ {\em constants},
which are the values of these functions in the solar neighbourhood. 
Thus, in practice they tell us about the local shape of the rotation
curve.  However, it is important to note that the range over which the
rotation curve parameters are effectively averaged depends on the range
in Galactic radii over which the stellar observations are made. 

The orbits of stars in the solar neighbourhood can be described as an
epicyclic motion due to their random motions with respect the local
standard of rest, superimposed on a circular orbit \cite{jhO65}.  The
length of the semi-major axis of the epicycle in the radial direction
($a_{epi}$) depends on the velocity in the radial direction ($\Pi$):

\begin{eqnarray}
a_{epi} &=& 0.5 \, \Pi / \sqrt{-B (A-B)}  \, ,
   \label{eqn:a_epi}
\end{eqnarray}

\noindent which equals $\sim$350, $\sim$700, and $\sim$1200 parsec for
Cepheids, early-type stars, and late-type stars, respectively.  Thus,
the random motions of stars introduces some form of radial smearing of
kinematical properties which precludes significant variation of
kinematical properties on scales smaller than $a_{epi}$.

Perhaps the greatest difficulty in interpreting the estimated values for
the Oort constants comes from the presence of the derivative terms in
equations (\ref{eqn:Oort_A}) and (\ref{eqn:Oort_B}): the values for the
Oort constants depend on any local bumps and wiggles in the rotation
curve as well as its global form.  Any attempt to interpret the Oort
constants must therefore consider the local topography of the rotation
curve in the solar neighbourhood.  Unfortunately, the available
observations of the Milky Way's rotational motion are not good enough to
allow one to calculate the derivatives in $\Theta(R)$ directly from the
data.  Instead, some suitable functional form must be fitted to the
data, and the derivatives of the model can then be calculated for
comparison with the Oort constants. 

The customary technique for fitting a model to rotation curve data for
an external galaxy is to decompose the galaxy into different mass
components: a stellar bulge and disk; an azimuthally-symmetric molecular
and atomic gas disk; and a dark matter halo
\cite{egAS86,smK87,kBeg89,LF89,ahB92,PaperI,PaperII}.  The free
parameters in this mass decomposition are then constrained by requiring
that the rotation curve for the model provides the best match possible
to the observed kinematic data.  In this paper, we seek to apply the
same basic technique to the kinematic data for the Milky Way in order to
produce a dynamical picture consistent with the observed Oort constants. 

Although the dominant contributors to the total mass come from the
stellar components and the dark halo, the contribution from the gas
cannot be neglected in this application.  While the stars and dark
matter are believed to be fairly smoothly distributed with radius, the
distribution of molecular and atomic hydrogen (\Ht \ and \HI
respectively) often show density enhancements like rings and arms.  Such
irregular distributions will produce a contribution to the total
rotation speed that can vary non-monotonically with radius.  Thus,
although these components do not contribute much to the amplitude of the
rotation curve, they can have a large effect on its local derivative,
and hence the predicted values of the Oort constants. 

We start by presenting the Milky Way's `observed' rotation curve and its
uncertainties in \S \ref{sec:The_MilkyWays_rotation_curve}.  In section
\ref{sec:The_HI_and_H2_distributions}, we calculate the radial
distribution of gas in the Milky Way, and show that it does, indeed,
significantly affect the analysis of the Oort constants.  Section
\ref{sec:RC_MM_OC} presents the mass component decomposition of the
Milky Way's rotation curve, and shows that the resulting models are
completely consistent with the constraints from the Oort constants as
long as we adopt values for $R_0$ and $\Theta_0$ at the lower end of the
acceptable range.  In section \ref{sec:Discussion}, we discuss the
implications of these results.  And we summarize in section
\ref{sec:Summary}.

\vspace*{-0mm}
\section{The Milky Way's rotation curve}
\label{sec:The_MilkyWays_rotation_curve}

\begin{figure*}
\begin{center}
 \epsffile{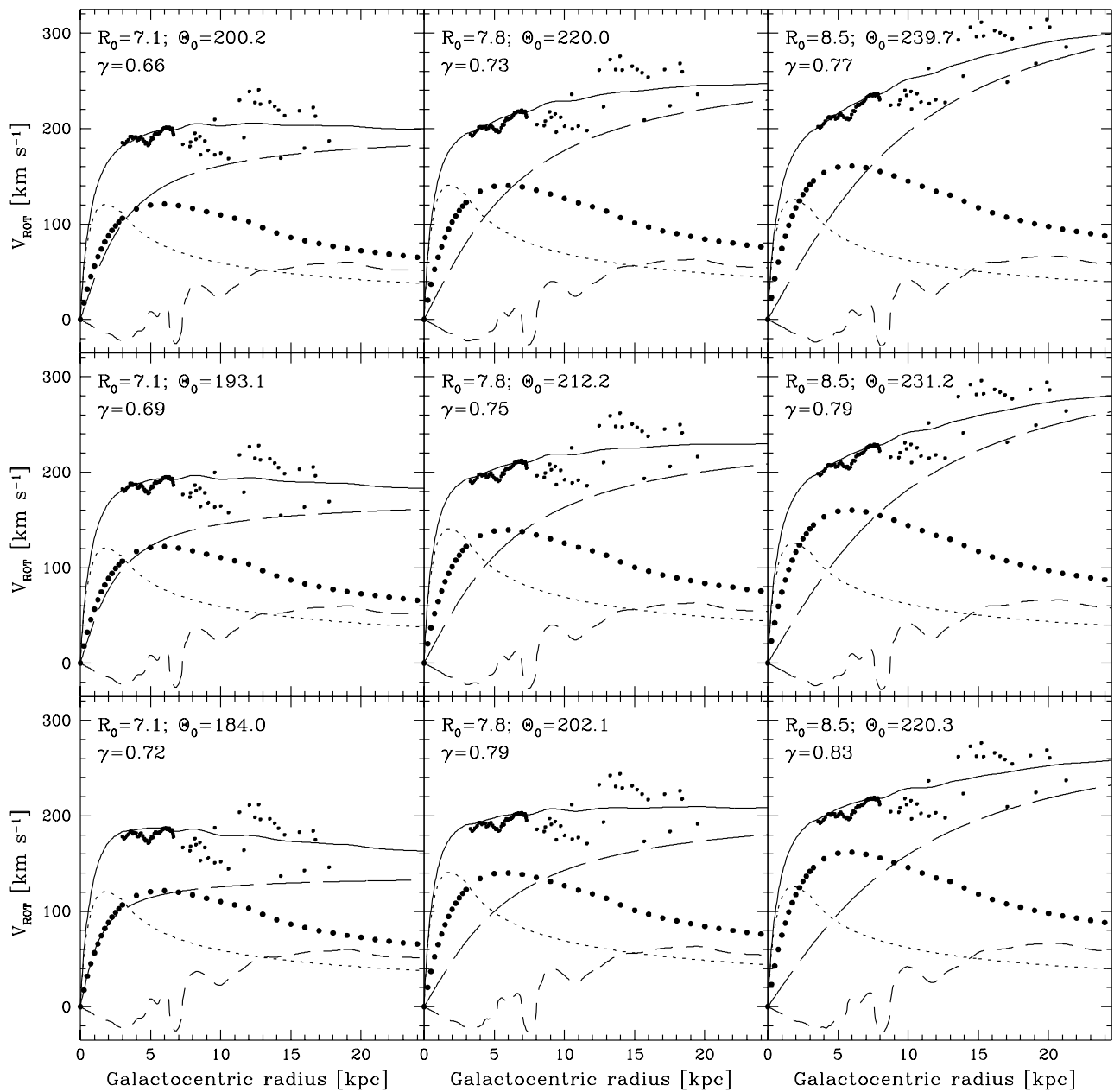}
 \caption{ \label{fig:GlobalRotCurves} Nine possible rotation curve
(dots) for the Milky Way which are consistent with the tangent point
data (Malhotra 1995), the radial velocities of \HII\ regions (Brand \&
Blitz 1993) and the \protect\HI\ $W(R)$ curve (Merrifield 1992).  For
clarity, we did not plot the error bars on the observed data points (a
few \protect\kms, and $\sim$ 15 \protect\kms in the inner and outer
Galaxy, respectively).  Each panel corresponds to a different choice for
\protect\RSUN and \protect\VSUNn, where all \protect\RSUNn-\protect\VSUN
combinations are consistent with the Kerr \& Lynden-Bell values at the
1.3-$\sigma$ level.  From left to right the Galactocentric distance
\protect\RSUN increases from 7.1 through 7.8 to 8.5 kpc.  In the bottom
row we present our best-fitting model (left panel; \protect\RSUN = 7.1
kpc, \protect\VSUN = 184 \protect\kms, A-B = 25.92 \protect\kms
kpc\protect\rtp{-1}; see \S \ref{sec:Discussion}), Sackett's suggested
values (middle panel; 1997), and Kerr \& Lynden-Bell's recommendations
(right panel; 1986).  From bottom to top $(A-B)$ increases from our best
fitting value of $(A-B)$ = 25.92 \protect\kms kpc\protect\rtp{-1},
through Feast \& Whitelock's (1997) best fit value $(A-B)_{FW} = 27.2$
\protect\kms kpc\protect\rtp{-1}), to $(A-B)$ = $(A-B)_{FW}$ +
1-$\sigma$ = 28.2 \protect\kms kpc\protect\rtp{-1}.  For smaller values
of $(A-B)$ and/or \protect\RSUNn, the rotation curve will decline even
more with radius.  The mass model fit to the rotation curve is shown as
a solid line (\S \ref{sec:Mass_models}).  Also shown are the
contributions from the individual components: the bulge (dotted line);
the stellar disk (filled circles); the dark halo (long dashed line); and
the gas layer (short dashed line).  Notice that the contribution of the
dark halo increases for smaller values of \protect\RSUN and larger
rotation speeds.  The degree to which the disk is maximal ($\gamma$;
Olling 1995) is indicated in each panel.  Note that the total {\em
stellar matter} \protect\cite{pdS97} contributes somewhat more:
$\sim$74\%, $\sim$81\%, and $\sim$81\% for \protect\RSUN = 7.1, 7.8, and
8.5 kpc, respectively.  The non-linear behaviour of the rotation curve
due to the gas in the Solar neighbourhood introduces a tiny wiggle in
the total model rotation curves, similar effects are seen around $R$ =
0.6\protect\RSUN and 1.4\protect\RSUNn. 
}

\end{center}
\end{figure*}

We have constructed the rotation curve of the inner Galaxy from the
tangent point measurements \cite{sanM95}.  For the outer Galaxy, we have
used determinations of the $W(R)$ curve [Eqn.  (\ref{eqn:WR})] based on
\Ht\ regions \cite{jBlB93} and the thickness of the gas layer
\cite{mrM92}.  As mentioned in the introduction, the rotation curve
inferred from radial velocity observations depends significantly on the
adopted values of \VSUN and \RSUNn, and so we constructed about 50
``observed'' rotation curves for 5 choices of \RSUN (6.1, 6.8, 7.1, 7.8,
and 8.5 kpc) and about 10 values of \VSUN between 170 and 230 \kms.  All
of these rotation curves are consistent with the observed radial
velocity measurements in the inner and outer Galaxy.  Generally
speaking, small values for \VSUN and/or \RSUN yield declining or flat
rotation curves, while the rotation curves constructed for large \VSUN
and/or \RSUN values tend rise in the outer Galaxy. 
Figure~\ref{fig:GlobalRotCurves} shows some typical models. 

A comparison of the Galactic rotation curve with rotation curves of
external galaxies is best made when the radial extend is re-scaled in
terms of the optical scale-length ($\hR = 2.5 \pm 0.5$ kpc; \S
\ref{sec:Mass_models}).  The last measured point on the Milky Way's
rotation curve lies at 2\RSUNn, so that the extent falls somewhere
between approximately 4.7\hR\ $(=2 \times 7.1 / 3)$ and 8.5\hR\ $(=2
\times 8.5 / 2)$.  It it not uncommon for galaxies to have rotation
curves which decline somewhat between a peak at around 2.3\hR\ and 5\hR. 
This is in particularly common in systems with large bulge-to-disk
ratios\footnote{The Milky Way's bulge-to-disk ratio is approximately
0.27, in K-band \cite{rpOmrM98a}} like M31 \cite{smK89}, NGC 2683 \& NGC
3521 \cite{sCjvG91}, and many others \cite{vRetal85,kgB87,ahB92}.  In
terms of the universal rotation curve \cite{PSS96}, 200 \kms\ is a
dividing point: on average, systems which rotate faster decline beyond
2.3\hR, while for slower rotators the trend is to have a rising rotation
curve.  From the comparison with external galaxies we conclude that
\VSUN = 180 \pmt 20 \kms, while a rotation speed at the solar circle as
high as 220 \kms\ is highly unlikely (see also Merrifield 1992).

The accuracy of the Milky Way's rotation curve -- reflected by the
scatter of the data points in Fig.~\ref{fig:GlobalRotCurves} -- is much
poorer than that routinely obtained for external galaxies.  This is
partly explained by the fact that for external galaxies a large range in
azimuth can be used, while for the Milky Way only two regions in
Galactocentric longitude around $l=135\degr$ and $l=225\degr$ [cf.  Eqn. 
(\ref{eqn:Vrad_apx})] are useful.  Only in those regions is the
$v_r$-distance gradient large enough to rise above the noise induced by
the tracer's random motions.  Furthermore, small errors in the tracers
distances or small non-circular motions in regions away from
$l=135\degr$ and $l=225\degr$ can result in large changes in the
inferred rotational velocities.  Clear examples of this effect are the
deviant points on our rotation curve between 1 and 1.2\RSUN which arise
from radial velocities of \Ht\ regions between $l=90\degr$ and
$l=120\degr$ \cite{jBlB93}.  These data points are likely to be affected
by non-circular motions \cite{wDjB97} induced by the nearby Perseus arm
\cite{jBlB93}, and distance errors resulting from unaccounted for
extinction, or errors in the photometric parallaxes (Oudmaijer, private
communications).  These large errors preclude the determination of the
Oort constants from the Milky Way's observed rotation curve.  As the
best alternative, we will model the rotation curve and determine the
Oort constants from the smooth model fits instead.  Rather than using
the customary linear or power-law fits to the rotation curve, we use the
mass modeling procedure which has proven highly successful in external
galaxies (\S \ref{sec:Mass_models}).

\begin{figure}
\begin{center}
 \epsffile{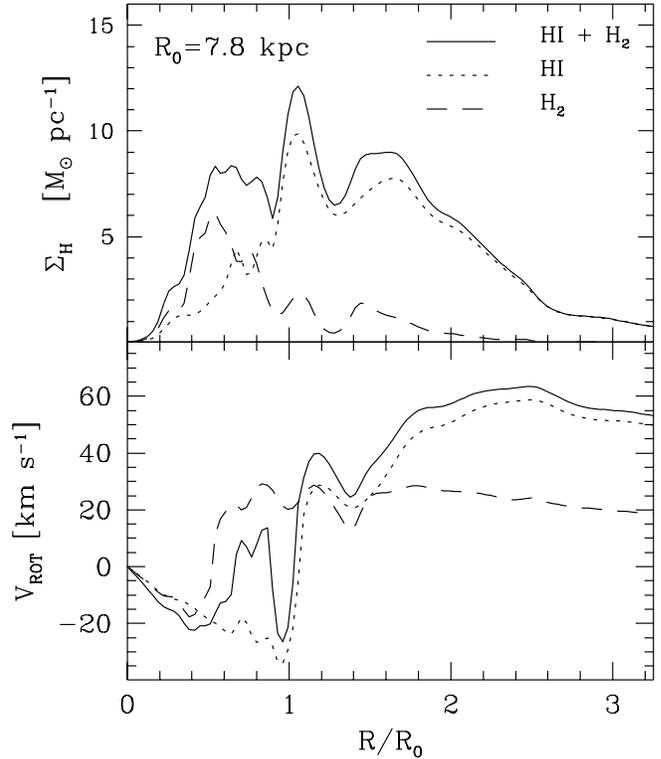}
 \caption{ \label{fig:HI_H2_columndensities} 
 The upper panel shows the face-on surface density distribution of the
\protect\HI (dotted line), the \protect\Ht (dashed line), and the sum of
two (full line) for the Milky Way.  The lower panel shows the
contribution these components make to the circular speed of the Galaxy,
where a negative rotational velocity means that the radial force is
directed outwards.  For other values of $R_0$, the surface density is
unaffected, but the rotation speeds vary approximately proportional to
$R_0$.  The radial force due to the ISM varies strongest in the regions
around 0.6\RSUNn, \RSUNn, and 1.4\RSUNn.
} 

\end{center}
\end{figure}

\vspace*{-0mm}
\section{The \protect\HI and \protect\Ht \ distributions}
\label{sec:The_HI_and_H2_distributions}

In order to determine the radial variation of the column density of
the molecular and atomic hydrogen, we need to be able to assign a
distance to a parcel of gas observed at Galactic coordinates $l$ and
$b$, and radial velocity $v_r$.  Assuming that the gas is on a
circular orbit, its distance $d$ can be determined using the following
relations:

\begin{eqnarray}
W(\frac{R}{R_0}) &=& \frac{v_r}{\sin{l} \cos{b}} =
   \frac{R_0}{R} \times \Theta(\frac{R}{R_0}) - \VSUN \, ,
   \label{eqn:WR} \\*[2mm]
R^2 &=& \RSUN^2 + d^2 \cos^2{b} - 2 \RSUN d \cos{l} \cos{b} \, .
   \label{eqn:Eqn.for.distance}
\end{eqnarray}

\noindent While the radial velocity and $W(\frac{R}{R_0})$ can be
determined accurately \cite{sanM95,mrM92,jBlB93}, the uncertainties in
\RSUN and \VSUN lead to a significant uncertainty in the value of $d$. 
We have therefore performed this analysis using a range of values for
\RSUN and \VSUNn.  Using the rotation curves which follow from the
adopted values for \RSUN and \VSUNn, we determined the volume and column
densities of the molecular and atomic gas.  For the inner Galaxy, the
\HI \ column density was determined from the midplane volume density
\cite{bB88} and the observed thickness of the layer \cite{sanM95}.  The
column densities for $R \ge R_0$ were taken from Wouterloot et al. 
(1990), with 9.25 \MSpcsq\ at the Solar circle.  The \Ht \ column
densities for the inner and outer Galaxy were taken from Bronfman et al. 
(1988) and Wouterloot et al.  (1990), respectively.  The column density
of \Ht\ at the solar circle is 1.8 \MSpcsq, and so, neglecting the
column density due to the other phases of the ISM, we arrive at a total
column density for the ISM at the solar circle of 14.5 \MSpcsq, which
includes 23.8\% Helium by mass \cite{kaOgS95}.  Note that while the
column density at fractional radius $R/\RSUN$ is independent of the
choice of \RSUN \cite{BCAMT88}, the Milky Way's total gas mass does
depend on this value.

\begin{table*}
 \caption{The Oort Constants as determined by several groups: KLB86
\protect\cite{KLB86}, PMB94 \protect\cite{PMB94}, BB93
\protect\cite{jBlB93}, and FW97 \protect\cite{mFpW97}.  We adopt the
values as determined by H87 \protect\cite{rbH87}.  The units of $A$,
$B$, $(A-B)$, and $(A+B)$ are km s$^{-1}$ kpc$^{-1}$, \protect\RSUN is
in kpc, and \protect\VSUN in \protect\kms.  The ratio $-B/(A-B)$ may be
equated with the square of the tangential to radial stellar velocity
dispersion ($X^2=\sigma_{\phi}^2/\sigma_R^2$).  Most stars of spectral
type G through K fall in the ``low $X^2$''.  M type stars fall mostly in
the ``high $X^2$'' category (see \S
\protect\ref{sec:Oort_constants_constraints}). 
}
 \label{tab:Oort_Constants}

\begin{tabular}{l|cc|c|c|c|c|c}
       &  $A$       &  $-B$        &$A-B=\Theta_0/R_0$&$A+B=(d\Theta/dR)|_{R_0}$     &$\frac{-10B}{A-B}$& $R_0$ & $\Theta_0$ \\ \hline
KLB86  &$14.5\pm 1.3$&$12.0\pm 2.8$&$26.4\pm 1.6$&$+2.5\pm 3.1$& $4.2 \pm 0.6$    & 8.5\pmt 1.1 & 220 \pmt 20  \\
BB93   &$12.6       $&$13.2       $&$25.9       $&$-0.6       $& $5.1$            & 8.5 & 220 \\
PMB94  &$15.9\pm 0.3$&             &             &             &                  & 8.1\pmt 0.3 & \\
FW97   &$14.8\pm 0.8$&$12.4\pm 0.6$&$27.2\pm 1.0$&$+2.4\pm 1.0$& $4.5 \pm 0.2$    & 8.5\pmt 0.5 & 231 \pmt 15\\
H87    &$11.3\pm 1.1$&$13.9\pm 0.9$&$25.2\pm 1.9$&$-2.6\pm 1.4$& $5.5 \pm 0.5$    &  &  \\
This work &$11.3\pm 1.1$&$13.9\pm 0.9$&$25.2\pm 1.9$&$-2.6\pm 1.4$&              & 7.1\pmt 0.4 & 184 \pmt 8 \\
low $X^2$&           &             &             &             & $3.5 \pm 0.5$    & & \\
high $X^2$&          &             &             &             & $5.0 \pm 0.5$    & & \\ \hline
\end{tabular}
\end{table*}

The \HI\ and \Ht\ column densities, and these mass components'
contributions to the total rotation curve, are presented in
Fig.~\ref{fig:HI_H2_columndensities}.  It is apparent that the sun
resides in a part of the Milky Way where the gradient of the gas column
density is substantial, and that their corresponding contributions to
the gradient in the rotation curve are large.  Clearly, this effect must
be taken into account when interpreting the Oort functions of equations
(\ref{eqn:Oort_A}) and (\ref{eqn:Oort_B}).

Note however that this effect is not limited to the Solar neighbourhood:
the regions around 0.6\RSUN and 1.4\RSUN exhibit similar behavior. 
Examples of such non-monotonic rotation curves and \HI surface density
distributions can be found in many external galaxies \cite{kgB87}.  Some
galaxies which have been observed with high linear resolution show
strong radial \HI\ density gradients: NGC 4244 \cite{PaperII}, NGC 891
\cite{jBfCfV97}, M31 \cite{eBbB84}, M33 \cite{eDjvdH87}, and NGC 1560
\cite{ahB92b} all show similar radial density gradients to those found
in the Milky Way.

\vspace*{-0mm}
\section{Rotation Curve, Mass models and the Oort constraints}
\label{sec:RC_MM_OC}

In order to produce a complete mass model for the Milky Way, we must fit
all the components  -- gas, stars and dark matter --  to the observed
rotation curve.

\vspace*{-0mm}
\subsection{Mass models}
\label{sec:Mass_models}

In a previous section, we obtained the gas mass distribution, so we now
turn to the stars and dark matter.  We have modeled the stellar
component as the sum of a bulge and disk, whose light distribution is
consistent with Kent's (1992) photometry, with the disk scalelength
forced to lie in the range \hR=2.5 \pmt 0.5 kpc \cite{pdS97}.  The dark
halo was modeled as a non-singular isothermal sphere.  We refer the
reader to a related paper \cite{rpOmrM98a} for a more detailed
description of the mass models. 

We then solved for the free parameters in this model by fitting the
observed rotation curve to the predictions of the mass model.  This
procedure that has been followed in many previous analyses
\cite{CO81,BSS83,RK88,kKgG89,smK92,mrM92,EJ94,Aetal95,GGT95,wDjB97},
but, to our knowledge, none of the previous studies have included the
actual radial surface density distribution of the Milky Way's ISM.  In
addition, we imposed the constraint that the mass model had to reproduce
the observed stellar surface density of the Milky Way in the solar
neighbourhood, $\Sigma_* = 35 \pm 5$ \MSpcsq\ (Kuijken \& Gilmore 1989). 

For each of the 50 \RSUNn-\VSUN combinations considered, we constructed
nine mass models with \hR=2, 2.5, and 3 kpc, and three stellar
mass-to-light ratios ($\Upsilon_*$) such that $\Sigma_*$ is recovered to
within \pmt 1-$\sigma$ \cite{rpOmrM98a}.  We show nine of the $\sim$450
mass models -- chosen to be representative of the ensemble -- in
Figure~\ref{fig:GlobalRotCurves}. 

\vspace*{-0mm}
\subsection{Oort constants constraints}
\label{sec:Oort_constants_constraints}

\begin{figure*}
\begin{center}
 \epsffile{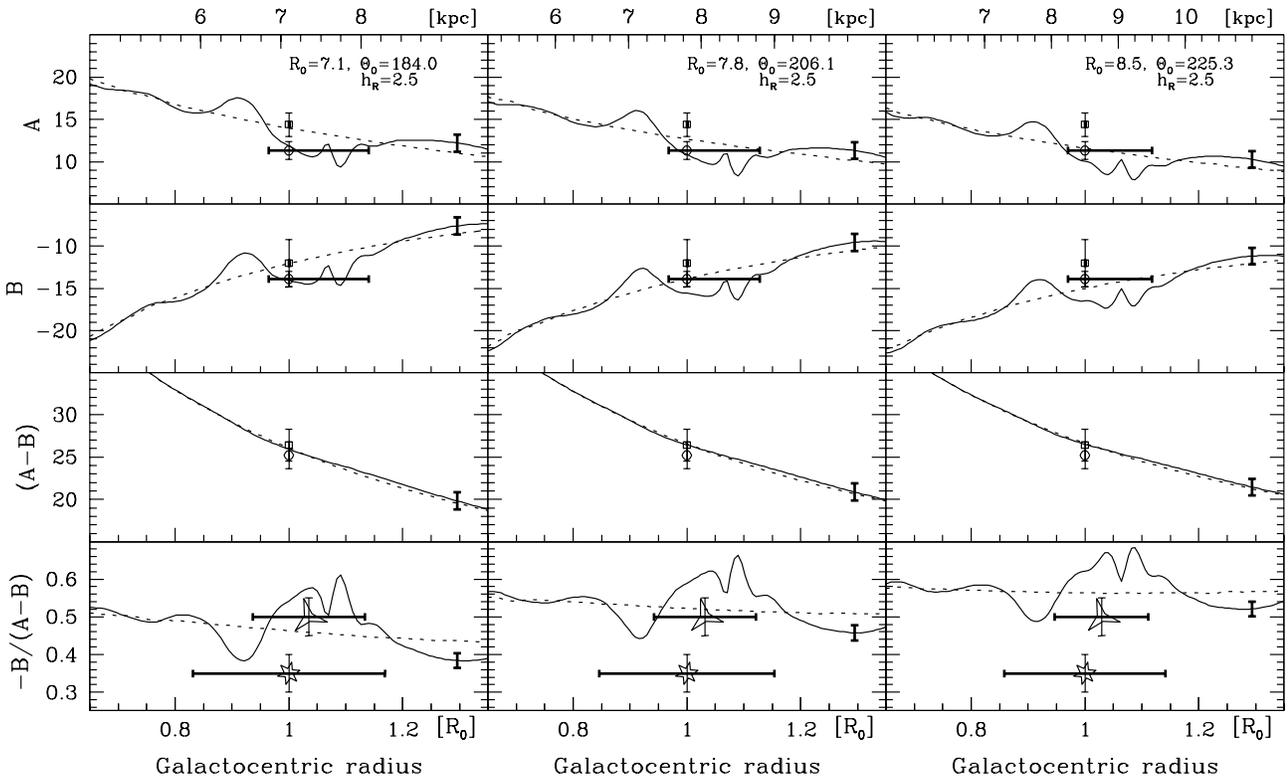}
 \caption{ \label{fig:Oort_Constants} The Oort functions, \OAR\ and
\OBR, as derived for three model rotation curves, together with the
values of $(A-B)$ and $-B/(A-B)$.  The units for $A$, $B$, and $(A-B)$
are ${\rm km \ s^{-1} \ kpc^{-1}}$.  The solid lines show the Oort
functions as derived from the full mass model, while the dashed line
shows what happens if the gas component is ignored.  The error bar at
the right of each panel shows the range of uncertainty due to variations
in the other model parameters ($\Upsilon_*$ and \protect\hR).  The
various observational estimates of these quantities are also shown, with
the horizontal error bars indicating the radial range over which the
observations effectively averaged (see text for details).  The upper
three panels show values as determined by KLB86 (open squares) and H87
(open circles).  The bottom panels show the values of $-B/(A-B)$ as
predicted by the velocity dispersions of ``high X$^2$'' stars (M dwarfs
and G-M giants; triangles) and ``low X$^2$'' stars (G-K dwarfs;
hexagons). 
}

\end{center}
\end{figure*}

From each mass model, we can calculate directly the Oort functions,
\OAR\ and \OBR, as defined by equations (\ref{eqn:Oort_A}) and
(\ref{eqn:Oort_B}).  Figure~\ref{fig:Oort_Constants} shows these
functions for models obtained from three combinations of \RSUN and
\VSUNn.  It is clear that the Oort functions (and the various
combinations thereof) vary significantly over even the limited range of
radii shown.  In the Solar neighbourhood about half of the variation in
\OAR\ and \OBR\ can be attributed to the general $\Theta(R)/R$ trend
(dotted lines), while the local deviations are caused by the
contribution of the gaseous component to the rotation curve.  It is also
apparent that the values of these functions vary significantly over the
acceptable range of values for \RSUN and \VSUNn.

The functions also change to some extent when we adopt different
acceptable combinations of the other parameters in the mass model
($h_R$, $\Upsilon_*$, and the parameters of the dark halo).  However,
these modifications are all small.  The error bar toward the right of
each panel in Figure~\ref{fig:Oort_Constants} indicates the full range
of variations induced by such changes.  Since the acceptable
combinations of stellar and dark matter components all reproduce the
same large-scale smooth variation with radius of the rotation curve, the
resulting Oort functions scarcely depend on the particular combination
adopted.  The dependence on \RSUN and \VSUN arises because changing
these quantities alters the largescale shape of $\Theta(R)$ (see
Fig.~\ref{fig:GlobalRotCurves}).  Thus, although the observed values for
various Oort constraints will not distinguish between the details of the
various degenerate mass models, they will allow us to decide which
combinations of \RSUN and \VSUN are consistent with the observed stellar
kinematics. 

The observed values for the various Oort constraints, taken from Kerr \&
Lynden-Bell (1986; hereafter referred to as KLB86) and Hanson (1987;
henceforth H87) are summarized in Table~\ref{tab:Oort_Constants}, and
are also plotted in Figure~\ref{fig:Oort_Constants}.  Since the Oort
functions vary substantially with radius, it is vital that we compare
the observed values to the functions over the appropriate range of
radii.  If we had not included the contribution from the gas component
(dotted lines in Figure~\ref{fig:Oort_Constants}), the Oort functions
\OAR\ and \OBR\ would have changed steadily with Galactocentric radius
[$dA/dR \sim-1.9, dB/dR \sim2.5$ (\kms kpc\rtp{-1})/kpc].  However, the
contribution of the gas flattens the \OA\ and \OB\ gradients
substantially over the scale of the width of the \HI\ crest which peaks
at 1.1\RSUNn.  In this radial regime we can thus use the small distance
approximations  -- which require constant \OA\ and \OB\ values --  for $v_r$
and $\mu_l$ [Eqns.  (\ref{eqn:Vrad_apx}) and (\ref{eqn:mu_l_apx})]. 

Unfortunately, KLB86 obtained their estimates by averaging together a
number of previous determinations, and it is not clear what range in
radii the data for these various analyses came from.  As KLB86 noted, it
is even unclear if the individual estimates were all obtained from data
at commensurate radii.  We are therefore unable to estimate the
appropriate radial range for these estimates.  H87 used a more
consistent set of data from the Lick Northern Proper Motion survey,
consisting of 60,000 stars within about 1 kpc of the Sun.  From their
$B-V$ colors [0.84 \pmt 0.5, \cite{KJH87}] we infer that these stars are
F2-K0 dwarf stars \cite{MB81} with a radial dispersion of roughly 29
\kms, so that the epicyclic radius for these stars is $\sim$780 pc [cf. 
Eqn.  (\ref{eqn:a_epi})].  The Galactic longitudes of these stars ($l
\approx 50$ -- $195$ degrees; $|b| \ge 10\degr$) indicate that they are
mostly located in the outer Galaxy: the horizontal bar on the H87 data
points in Figure~\ref{fig:Oort_Constants} represents the range of $R$ of
a typical star in this sample [($R_0-250,R_0+1000)$ parsec].  Since this
region roughly equals the area which is covered by the epicyclic motions
of these F2-K0 dwarfs, they can be used to determine intrinsic Galactic
properties on this scale.  Furthermore, as discussed above, Hanson's use
of the approximate $\mu_l$ relation [Eqn.  (\ref{eqn:mu_l_apx})] is
justified due to the flatness of the \OAR\ and \OBR\ functions in this
region.

From Figure~\ref{fig:Oort_Constants} it is clear that the values of $A$
and $B$ derived by H87 are consistent with the Oort functions for small
values of \RSUN and \VSUNn.  Models without the contribution from the
gas component yield larger values for \OA\ and \OB.  It is also
noteworthy that the value of $2 A \RSUN$ based on determinations of
$W(R)$ in the inner Galaxy is 1.4 \pmt 0.25 times larger than the value
determined in the outer Galaxy \cite{mrM92}; this finding is also
entirely consistent with the Oort $A$-function curves shown in
Figure~\ref{fig:Oort_Constants}. 

An independent estimate for $(A-B)$ comes from proper motion surveys in
the directions $l = 90$ degrees and $l = 270$ degrees.  Objects in these
directions lie at radii close to $R_0$ out to quite large distances, and
so no radial range is plotted for the KLB86 and H87 estimates for this
quantity in Figure~\ref{fig:Oort_Constants}.  In any case, the absence
of a $d\Theta/d R$ term in this combination of Oort functions means that
$(A-B)$ is a smooth function of radius, so the importance of knowing the
precise radial range of the determination is reduced.  Once again, the
H87 value for $(A-B)$ is best matched by models with small values of
$R_0$ and $\Theta_0$. 

Finally, we turn to the observed value of $-B/(A-B)$.  If the potential
is azimuthally symmetric and the velocity dispersions are small, then
this combination of the Oort parameters is equal to the square of the
ratio of the tangential and radial stellar velocity dispersions,
$(\sigma_{\phi}/\sigma_{\rm R})^2$ \cite{kKsT94}.  The accepted value
for this ratio is $X^2=(\sigma_{\phi}/\sigma_{\rm R})^2$=0.42 \pmt 0.06
\cite{KLB86}.  However, upon close examination there appears to be a
bimodal distribution in $X^2$: dwarfs of spectral type G-K have a low
value [$X_{\rm D,G-K}^2=0.36 \pm 0.07$), while M dwarfs and G-M giant
stars have a somewhat larger dispersion ratio ($X_{\rm D,M}^2=0.5 \pm
0.11$, and $X_{\rm G,G-M}^2=0.5 \pm 0.07$; \cite{jD651,MB81}].  In fact,
the large error on $X^2_{\rm D,M}$ arises as a result of the bimodality
in the distribution of $X^2$ \cite{RHG97X2}, which is also clearly seen
in the distribution of space motions of individual stars
\cite{VJ51,jD65,jD65fDISP,MB81fDISP}.  Furthermore, these figures show
that the giant stars have a bimodal distribution as well.  This result
was confirmed by more recent investigations \cite{mM74,eO83}, which
found a minimum dispersion ratio for stars $\sim 1$ Gyrs old ($X^2 \sim
0.36$) while the dispersion ratios for older and younger stars are
larger ($X^2 \sim 0.46$).  Thus, rather than having to account for the
average $X^2$-value of 0.42 \pmt 0.06, we are faced with the task to
explain the bimodal $X^2$ distribution.  We estimate the location of the
two peaks to lie at: $X^2=0.35 \pm 0.05$, and $X^2=0.5 \pm 0.05$. 
Figure~\ref{fig:Oort_Constants} suggests that these results can be
explained if the bimodality in the dispersion-ratio measurements is a
result of the ``dip'' and the ``hump'' in the $-B/(A-B)$ curve around
0.9\RSUN and 1.1\RSUNn, respectively.  In this scenario, the dwarf
sample is dominated by stars originating within the Solar circle (small
$X^2$ values) with a smaller number of dwarf showing the larger
$X^2$-value, while the giants used in the dispersion measurements come
preferentially from the outer Galaxy.  Since the stellar density
increases towards smaller $R$ with a scale-length that may be as small
as 2 kpc, and since the observed distribution of giants is concentrated
around $(R-R_0)=250$ pc \cite{swM65}, this may well be the case.  The
larger range in radii covered by the dwarf stars with their larger
epicyclic excursions (indicated by the horizontal bars on the points in
Fig.~\ref{fig:Oort_Constants}) also means that these stars are more
influenced by the dip in the $-B/(A-B)$ curve at $R \approx 0.9 R_0$,
explaining the lower observed average value.  Once again, only models
with low values of $R_0$ and $\Theta_0$ are consistent with the
observations.

\vspace*{-0mm}
\section{Discussion}
\label{sec:Discussion}

\begin{figure}
\begin{center}
 \epsffile{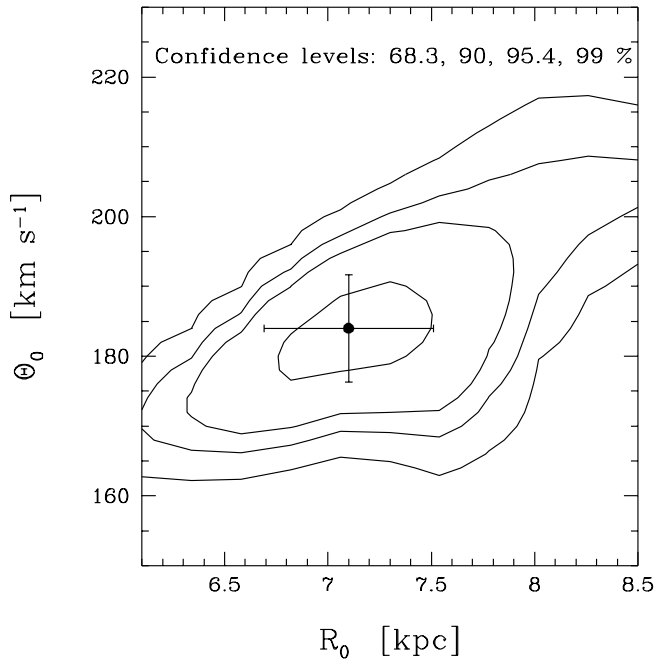}
 \caption{ \label{fig:R0_Theta0} The
contours of equal likelihood as a function of \protect\RSUN
and \protect\VSUN, calculated by comparing stellar kinematic
constraints to the model-predicted Oort constants.  The best-fit
minimum-$\chi^2$ values for \RSUN, \VSUNn, and their 1-$\sigma$ errors
are also indicated.}

\end{center}
\end{figure}

It is apparent from the above analysis that the local topography of
the rotation curve induced by the small-scale structure in the gas
component has a significant impact on the interpretation of the
stellar-kinematic Oort constraints.  Once this phenomenon is taken
into account, it is possible to obtain model rotation curves for the
Milky Way that are consistent with essentially all the Oort constraints.

It is also clear from Figure~\ref{fig:Oort_Constants} that a consistent
picture only emerges for certain values of $R_0$ and $\Theta_0$.  We can
formalize the constraints that this analysis puts on the values of the
Galactic constants by calculating a $\chi^2$ statistic comparing the
observed combinations of the Oort constants in
Figure~\ref{fig:Oort_Constants} to the values predicted by the models. 
Since this comparison requires that we know the radial range over which
the observations have been made, we use the observations plotted in
Figure~\ref{fig:Oort_Constants} for which we have an estimate of this
range (see above), together with the H87 estimate for $(A-B)$.

\begin{figure*}
\begin{center}
 \epsffile{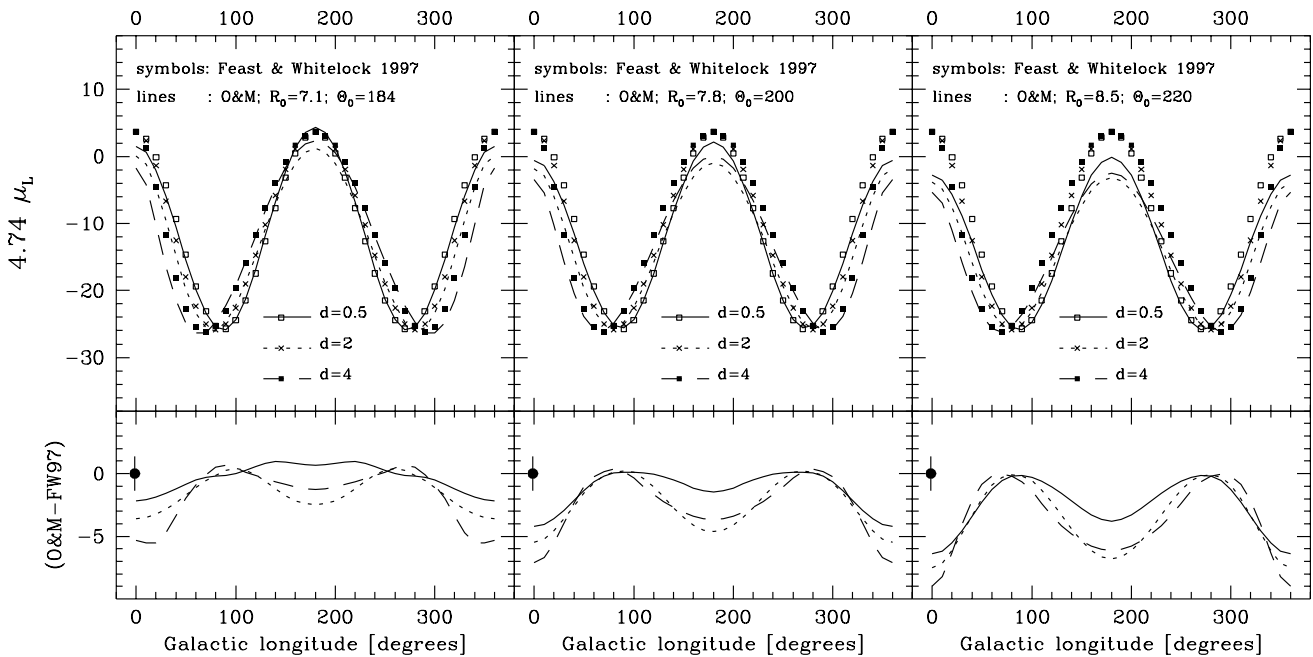}
 \caption{ \label{fig:Proper_Motions} Proper motions (top row) as
calculated using the Oort and Galactic constants derived by FW97
($R_0=8.5$ kpc, $A=14.82, B=-12.37$ km s$^{-1}$, and $d^2 \Theta / d R^2
= d^3 \Theta / d R^3 = 0$), and some of our models.  In the lower panels
we display the difference between the proper motions predicted by our
models and those predicted by FW97.  The units for the vertical axes are
\kms kpc\rtp{-1}.  All curves were calculated for $b=0$.  The error bar
in the bottom panel equals the quadratic sum of the errors of FW97's
determinations of \OA\ and \OB.  In the leftmost panel we present our
model which best fits the Oort constants constraints ($R_0=7.1$ kpc,
$\Theta_0=184$ km s$^{-1}$), while the other two panels were calculated
for the Sackett (1997; middle panels) and Kerr \& Lynden-Bell (1986;
right columns) recommendations for the Galactic constants.  Again, worse
correspondence between models and ``data'' occurs if ($|R_0-7.1|$)
increases.  In our models, assuming a different value for \VSUN will
shift the $\mu_l$ curves slightly up or down by $\Delta \Theta_0 / R$. 
}

\end{center}
\end{figure*}

The $\chi^2$ statistic comparing these five observations to suitably
radially-averaged models were then calculated for a range of values
for $R_0$ and $\Theta_0$.  The best-fit (minimum-$\chi^2$) values for
these parameters are $R_0 = 7.1 \pm 0.4$ kpc, and $\Theta_0 = 184 \pm
8$ \kms.  The complete two-dimensional confidence regions are plotted
in Figure~\ref{fig:R0_Theta0}.  The confidence levels measure the
probability that any given values for $R_0$ and $\Theta_0$ are
consistent with the observed Oort constraints.  Thus, for example, the
official IAU-sanctioned values of $R_0 = 8.5$ kpc and $\Theta_0 =
220$ \kms are ruled out at the 99\% confidence level.  It is,
however, notable that a value for \RSUN as low as 7.1 kpc is entirely
consistent with the only {\em direct} distance determination to the
Galactic center employing proper motions of H$_2$O masers, \RSUN = 7.2
\pmt 0.7 kpc \cite{mjR93}.

The proper motion of Sgr A$^*$, $\mu_{Sgr A^*}$
\cite{dcBraS87,dcB96,mjR98}, has been used to used to constrain
$(A-B)$ as well \cite{aGsvR97}.  These quantities are related via the
formula,
\begin{eqnarray}
\mu_{Sgr A^*} \hspace*{-3mm} &=& \hspace*{-3mm}
       -\frac{A-B}{4.74} - \frac{v_0 + v_{Sgr A^*}}{4.74 \RSUN}\,{\rm
milliarcsec}\,{\rm year}^{-1},
   \label{eqn:mu_Sgr}         
\end{eqnarray}
where $v_0$ is the Solar motion relative to the local standard of rest
($v_0 = 5.3 \pm 1.7$ \kms; Binney et al.\ 1997) and $v_{Sgr A^*}$ is
the velocity of Sgr A$^*$ with respect to the Galactic centre ($v_{Sgr
A^*} \le 11$ \kms; Gould \& Ramirez 1997).  Using the range of
acceptable values for $(A - B)$ derived from our mass models, we find
that the proper motion of Sgr A$^*$ should lie in the range $\mu_{Sgr
A^*} = -5.95 \pm 0.36$ milliarcsec per year, which is entirely
consistent with the observational constraints.

\vspace*{-0mm}
\subsection{Other determinations of the Galactic constants}
\label{sec:Other_determinations_of_the_Galactic_constants}

Inspection of Table~\ref{tab:Oort_Constants} reveals that the Oort
constants as derived by different authors can vary by up to
4.2-$\sigma$.  We have used the determination of the Oort constants
which are based on ground based proper motions of relatively nearby
stars ($d\le 1$ kpc; H87) in the outer Galaxy.  Cepheids probe a much
larger range in distance and can be used to determine the Oort constants
as well, but yield significantly different values
\cite{PMB94here,mFpW97here}.

Let us first examine the region within 1 kiloparsec from the Sun where
the non-linearity of the Oort functions is most prominent.  As discussed
above, the H87 determined \OA\ and \OB\ in the ``flat'' region extending
from 0.95\RSUN to 1.15\RSUNn: $A_{H87}$ = 11.3 \pmt 1.1 \kms
kpc\rtp{-1}.  The Cepheids on the other hand are located preferentially
in the inner Galaxy (PMB94, their Fig.  4) where our models yield an average
\OA\ of 15.4 \pmt 2, entirely consistent with their value of 15.9 \pmt
0.3.  From Figure~\ref{fig:Oort_Constants} it is also clear that beyond
1 kpc the non-linearities in \OAR\ and \OBR\ are much smaller.  In order
to assess whether the PMB94 \& FW97 results are consistent with our
models at large distances, we have compared the radial velocities and
proper motions as predicted by the PMB94 \& FW97 models with our models. 
Since the PMB94 \& FW97 models are fitted directly to their data, we are
in effect comparing our models with their data.  Our model rotation
curves yield exact radial velocities with equation (\ref{eqn:WR}), and
exact proper motions using the formula

\begin{eqnarray}
\mu_l \hspace*{-3mm} &=& \hspace*{-3mm}
   \left( \frac{\Theta(R) \left( \RSUN \cos{l} - d \right) \cos{b} }{R} -
          \VSUN \cos{l} 
   \right) \hspace*{-1mm} \frac{1}{4.74 d} \, ,
   \label{eqn:mu_l}         
\end{eqnarray}

\noindent in units of milli arcsec per year, with the distances in units
of kiloparsec \cite{MB81}.  In the upper panels of
Fig.~\ref{fig:Proper_Motions} we present the proper motion curves as
modeled by FW97 (symbols) and some of our models (lines) for a range
of values for \RSUNn, \VSUNn, and $d$.  In the bottom panels we plot
the difference between the two.  Our model which best fits the Oort
constant constraints (\RSUNn=7.1 kpc, \VSUNn=184 \kms) is displayed in
the leftmost column; the agreement between the FW97 proper motions and
our predictions is excellent.  The worst agreement between data and
model is found for large distances towards the Galactic centre, where
the rotation curve is rather ill-defined as a result of bar-induced
non-circular motions.  Furthermore, we note that models with larger
and smaller values for \RSUN fit the FW97 data progressively worse.
We draw the same conclusion from a similar comparison of the PMB94
radial velocity data with our models.

From Figure~\ref{fig:Proper_Motions}, we can also see that the vertical
scatter seen in observational $l-\mu_l$ plots is not due to some
intrinsic scatter, but rather results from the fact that the data occupy
a finite range in radii.  Clearly the approach followed by PMB94 \& FW97
-- fitting model $v_r$ and $\mu_l$ curves which include distance effects
-- is superior to the application of the approximate equations [Eqns. 
(\ref{eqn:Vrad_apx}) and (\ref{eqn:mu_l_apx})].

Given that the best-fit values of the Galactic \& Oort constants in this
analysis are so different from the Cepheid derived values it behooves us
to explain why our best model nevertheless is able to fit the radial
velocity and proper motion data (Fig.~\ref{fig:Proper_Motions}) so well. 
The reason for this is that the Milky Way's rotation curve and the
Galactic \& Oort constants are secondary parameters: a wide range in
secondary parameters is consistent with the primary observable
$W(\frac{R}{R_0})$ (Fig.~\ref{fig:GlobalRotCurves}).  So assuming that
the $W(\frac{R}{R_0})$ curves as measured by Brand \& Blitz (1993),
Merrifield (1992), and Pont, Mayor \& Burki (1994) are a fair sample of
the intrinsic Galactic velocity field, it is not surprising that our
smooth model fits to the first two data sets are consistent with PMB94's
Cepheids data.  Varying the Galactic constants has only a small effect
on the proper motions, as can be seen by rewriting equation
(\ref{eqn:mu_l}) in terms of the primary observable $W(\frac{R}{R_0})$:

\begin{eqnarray}
\mu_l \hspace*{-3mm} &=& \hspace*{-3mm}
   \frac{1}{4.74}
   \left(
      \frac{W(\frac{R}{R_0}) \cos{l}}{d} - 
      \frac{W(\frac{R}{R_0})}{\RSUN} - (A-B)
   \right) \, .
   \label{eqn:mu_l_WR}
\end{eqnarray}

\noindent Changing $(A-B)$ shifts the $\mu_l$ curves up and down, but
the observed error on $(A-B)$ corresponds to a small vertical shift of
$\mu_l$, which is only significant in the Galactic centre and
anti-centre regions.  Likewise, changing \RSUN affects the ratio
$\frac{R}{R_0}$ -- and hence $W$ -- only slightly [through Eqn. 
(\ref{eqn:Eqn.for.distance})].  Like a change in $(A-B)$, adjusting
\RSUN shifts the $\mu_l$ curves up and down.  The only longitudinal
dependence comes from the first term, which can be large for small
distances $d$.  Thus, the differences between the proper motion curves
for different choices of \RSUN \& \VSUN in Fig.~\ref{fig:Proper_Motions}
are not very large.  However, the residuals {\em do} increase if either
\RSUN or \VSUN or both are adjusted: our preferred model fits the proper
motion data best. 

\vspace*{-0mm}
\subsection{Uncertainties in the $\bmath{W(R)}$ curve}
\label{sec:Uncertainties_in_W_of_R}

The derived form for the Galactic rotation curve depends on the
adopted $W(R)$ curve.  In this paper, we have based our analysis on
the radial velocities of \HII\ regions (Brand \& Blitz 1993) and the
\protect\HI\ $W(R)$ curve (Merrifield 1992).  However, one could also
infer this curve from radial velocities of Cepheids.  As Dehnen \&
Binney (1997) have noted, the $W(R)$ curve derived from Cepheids drops
significantly more rapidly with $R$ than that obtained from the data
sets exploited here.  For such a rapidly-falling form for $W(R)$ to
produce a reasonably flat rotation curve, one must adopt a large value
of $\Theta_0$.  It is, therefore, unsurprising that Metzger, Caldwell
\& Schechter's (1998) analysis of Cepheid kinematics results in models
of the Milky Way with $\Theta_0$ as high as $240 \pm 10\kms$.  Their
modeling obtained a value for $R_0$ of $7.7 \pm 0.3\,{\rm kpc}$; the
corresponding value for $(A-B) = \Theta_0/R_0$ is $31 \pm 2\kms\,{\rm
kpc}^{-1}$, which is inconsistent with the various stellar kinematic
constraints summarized in Table~\ref{tab:Oort_Constants}.

Clearly, these internal inconsistencies mean that there must be
systematic errors in at least some of the available data.  The unknown
nature of these errors means that there can be no objective criterion
for choosing between the various methods.  The only way that a final
answer can be reached will be from the analyses of other kinematic
tracers.  Through such studies, one might hope to identify and prune
the discrepant data sets, obtaining a consistent picture from the
remaining majority.  With only three ``reliable'' techniques for
obtaining $W(R)$, we are not yet in a position to carry out such a
democratic procedure.

\vspace*{-0mm}
\subsection{The effects of non-circular motions}
\label{sec:The_effects_of_non_circular_motions}

One potential source of error in this analysis comes from our
assumption that the orbital structure of the Milky Way has an
azimuthal symmetry.  The presence of spiral structure means that this
assumption is almost certainly an over-simplification.  Non-circular
motions will have two effects on the above analysis.

First, in Section~\ref{sec:The_HI_and_H2_distributions} we used
kinematic distances to derive the distribution of gas.  These distance
assume that the gas follows circular orbits, and so the inferred
distribution will be distorted if there is significant non-circular
streaming.  However, dynamical models of spiral structure show that the
locations of high density ridges are quite accurately reproduced by the
naive kinematic modelling, and so the radial positions of the density
peaks and troughs apparent in Figure~\ref{fig:HI_H2_columndensities}
should be approximately correct, and the conclusion that we are located
in a region where this density varies rapidly with radius still holds. 
Second, the non-circular motions hamper the determination of the
intrinsic $W(R)$ curve from which the rotation curve, the Oort
constants, and the Milky Way's mass models are derived.  With
sufficiently large azimuthal coverage and dense sampling, such regions
can be identified [e.g., the Perseus arm between $l=90\degr$ and
$l=120\degr$, \cite{jBlB93}], and discarded.  However note that when
streaming motions have maximal effect on the observed radial velocities,
the effects on the proper motions will be minimal, and vice-versa. 
Thus, by combining the information from the line of sight and transverse
motions, all azimuths can be properly sampled, which should lead to a
significantly more accurate determination of the Milky Way's rotation
curve.  One could then use the deviations from the general axisymmetric
velocity field to study the non-circular motions in great detail and
infer the mass associated with the spiral arms. 

We have seen that the distribution of the ISM has a large effect on the
determination of the Oort constants in the Solar neighbourhood. 
However, the non-circular streaming motions have not been mapped out
within one kpc from the Sun because random motions dominate the radial
velocity and proper motion measurements of individual tracer objects. 
Although Hanson (1987) overcame this problem by using over 60,000 stars
in his determination of the Oort constants, his data set cannot be used
to investigate non-circular motions in the Solar neighbourhood as no
distances to individual objects are available for this survey. 

We should keep in mind that if streaming motions are present to a
significant degree, the stellar orbits will become more complex, and so
the Oort constraints derived from simple stellar kinematic measurements
will also be compromised.  Clearly, in order to allow for these
phenomena fully, one must compare a complete dynamical model for the
Milky Way to the observational data.  However, application of Occam's
razor to this problem suggests that the first step should be the
comparison of observations to the simpler azimuthally-symmetric model. 
The fact that we can satisfactorily describe the Oort constraints with
this model implies that there is no direct evidence for non-circular
motions in Hanson's Solar neighbourhood data.  If, however, future
observations establish that the true values for the Galactic constants
lie outside the acceptable region in Figure~\ref{fig:R0_Theta0}, then at
least one of the assumptions underlying our models must be wrong.  For
example, it might be that significant non-circular stellar motions are
present.  Alternatively, the mass distributions of the stars, or indeed
the dark matter may be non-smooth and contribute significantly to local
gradients in the rotation curve. 

\vspace*{-0mm}
\section{Summary}
\label{sec:Summary}

We have used distance and radial velocity data available in the
literature to investigate the rotation curve of the Milky Way.  The
accepted uncertainties in the Galactic constants of $\sim$10\% allow
for a wide range in rotation curve shapes.  We have fitted mass models
to about 450 rotation curves which are consistent with the radial
velocity data and determined the Oort functions \OAR\ and \OBR\ for
these model rotation curves.  Typically, the Oort functions vary at a
rate of a few \kms kpc\rtp{-2}.  In several locations in the Milky
Way, the more objective model rotation curves show sufficient fine
structure to affect the Oort functions significantly.  The ring of
\HI\ just beyond the Solar circle is massive enough to flatten \OAR\
and \OBR\ in the first kiloparsec beyond the Solar circle.  Comparison
between the model \OAR\ and \OBR\ functions and Hanson's (1987)
determinations of the Oort constants places tight limits on the
distance to the Galactic center and the Milky Way's rotation speed at
the position of the Sun: \RSUN = 7.1 \pmt 0.4 kpc, and \VSUN = 184
\pmt 8 \kms.  These results are consistent with the only direct
determination of the distance to the Galactic center, \RSUN = 7.2 \pmt
0.7 kpc, by Reid (1993), and the proper motion of Sgr A$^*$
\cite{dcBraS87,dcB96,mjR98}.  With these Galactic constants we find
that the rotation curve of the Milky Way declines slowly in the outer
Galaxy.  At larger distances from the Sun these models are entirely
consistent with recent determinations of the Oort constants based on
radial velocities and proper motions of Cepheids.

\vspace*{-0mm}
\section*{acknowledgements}

We thank Andy Newsam for useful discussions.  We also thank the
referee, Gerry Gilmore, for providing valuable suggestions, which have
greatly improved the manuscript.  MRM is supported by a PPARC Advanced
Fellowship (B/94/AF/1840).


\end{document}